\documentclass[aps,prl,
reprint
]{revtex4-1} 

\usepackage[utf8]{inputenc}

\usepackage{mathtools}
\usepackage{amssymb}
\usepackage{bm}
\usepackage{physics}
\usepackage[binary-units=true]{siunitx}

\usepackage{enumitem}
\usepackage[nameinlink,capitalise,noabbrev]{cleveref}
\crefformat{equation}{(#2#1#3)}
\Crefformat{equation}{Equation (#2#1#3)}
\crefrangeformat{equation}{(#3#1#4) to~(#5#2#6)}
\crefmultiformat{equation}{(#2#1#3)}%
{ and~(#2#1#3)}{, (#2#1#3)}{ and~(#2#1#3)}
\Crefmultiformat{equation}{Equations (#2#1#3)}%
{ and~(#2#1#3)}{, (#2#1#3)}{ and~(#2#1#3)}

\usepackage{graphicx}
\graphicspath{{./figures/}}

\usepackage[dvipsnames]{xcolor}

\definecolor{h}{rgb}{0 0 1}

\usepackage{psfrag}

\usepackage{pgf,tikz,pgfplots}
\pgfplotsset{compat=1.10}
\usetikzlibrary{shapes,arrows,calc,arrows.meta,
patterns,backgrounds,positioning,fit}
\tikzset{every picture/.style=black}
\usepackage{circuitikz}

\makeatletter
\newcommand{\getfontsize}{\f@size pt}
\makeatother

\newcommand*\ds{\displaystyle}

\newcommand{\avg}[1]{\left< #1 \right>}
\newcommand*\mean[1]{\avg{#1}}

\newcommand*\kB{k}
\newcommand*\qe{q_{\si{\elementarycharge}}} 

\newcommand*\Vsat{V_{\mathrm{sat}}}

\definecolor{k}{rgb}{0 0 0}
\definecolor{r}{rgb}{1 0 0}
\definecolor{g}{rgb}{0 1 0}
\definecolor{b}{rgb}{0 0 1}
\definecolor{orange}{rgb}{1,0.7,0}
\definecolor{c}{rgb}{0 1 1}
\definecolor{cc}{RGB}{64 224 208}
\definecolor{m}{rgb}{1 0 1}
\definecolor{khaki}{RGB}{128 128 0}
\definecolor{deepskyblue}{RGB}{0 191 255}
\definecolor{darkMagenta}{rgb}{0.5 0 0.5}
\definecolor{chocolateBrown}{RGB}{98 52 18}
\definecolor{lightBrown}{RGB}{189 154 122}
\definecolor{mybrown}{RGB}{127 37 0}
\definecolor{bordeaux}{RGB}{131 41 85}
\definecolor{violet}{RGB}{127 0 255}
\definecolor{myGreen}{RGB}{134,180,44}
\definecolor{gray_gate}{RGB}{211,208,205}
\definecolor{yellow_oxide}{RGB}{244,231,164}
\definecolor{color_mix}{rgb}{0.7510 0.2510 0.2510}

\definecolor{h}{rgb}{0 0 1}

\definecolor{l}{rgb}{0 0 0}

\newcommand*\Dsigma{\Delta \sigma}
\newcommand*\Dq{\Delta q}
\newcommand*\Dt{\Delta t}

\renewcommand{\avg}[1]{\left\langle #1 \right\rangle}
\renewcommand*\mean[1]{\avg{#1}}

\newcommand{\kBB}{k_B}

\begin{document}

\title{Moments of Entropy Production 
in
Dissipative Devices}

\author{Jean-Charles Delvenne}
\author{L\'eopold Van Brandt}
\affiliation{ICTEAM Institute, UCLouvain, Louvain-la-Neuve, Belgium
}

\date{\today}

\begin{abstract}
We characterize the possible moments of entropy production for general overdamped Markovian systems. We find a general formulation of the problem, and derive a new necessary condition between the second and third moment. We determine all possible first, second and third moments  of entropy production for a white noise process. 
As a consequence, we obtain a lower bound for the skewness of the current fluctuations in dissipative devices such as transistors, thereby demonstrating that the Gaussianity assumption widely used in electronic engineering is thermodynamically inconsistent.






\end{abstract}

\maketitle


\emph{Introduction.}---Stochastic thermodynamics extends and generalizes the laws of conventional thermodynamics and equilibrium statistical physics to   mesoscopic systems in which random fluctuations are non negligible~\cite{Peliti2021,Esposito2012,van2013stochastic}.
The theory is able to describe possibly strongly nonlinear systems operating far from equilibrium~\cite{Rao2016}. 
Modern nanoscale electronic devices, operating either in classical~\cite{Gabelli2009,IFAC2023,ICNF2023,Freitas2021stochastic} or quantum regime~\cite{Reulet2010,Maisi2014,Fevrier2019,Esposito2012,Gao2021,Freitas2020stochastic}, constitute one recent field of application of this theory. 
Recent contributions were dedicated to reliability assessment~\cite{Freitas2022reliability,EDTM2024}, or relation between noise and energy dissipation in digital CMOS circuits in non-stationary conditions~\cite{APL2023,wolpert2020thermodynamics,yoshino2023thermodynamics}.

The \emph{local detailed balance (LDB) relation}~\cite{maes2021local,van2013stochastic} results from the first principle of microscopic reversibility, and formulates the entropy production in terms of probabilities of direct and time-reversed trajectories. It holds in wide range of situations~\cite{Peliti2021}.
From the LDB one can derive a host of important results, such as the \emph{fluctuation relation}~\cite{Jarzynski1997}, or the \emph{thermodynamic uncertainty relations} (TUR)~\cite{Horowitz2020,barato2015thermodynamic,pietzonka2016universal,pietzonka2016current,polettini2016tightening,horowitz2017proof,proesmans2017discrete}.
The TURs provide a fundamental lower bound for the variance of entropy production $\Dsigma$, and more generally observables antisymmetric under time reversal~\cite{delvenne2024thermokinetic,Horowitz2020,Wampler2021,Salazar2022}.

Beyond mean and variance, characterization of higher-order moments, like the \emph{skewness} (the third central moment) 
quantifying
the asymmetry of the fluctuations, provides finer information about the random physical process~\cite{Pinsolle2018}, especially far from equilibrium~\cite{Salazar2022,IFAC2023}.
The topic is covered to a much lesser extent.
Other theoretical works focus on special cases, notably noninteracting systems~\cite{Ptaszynski2022}, unicyclic~\cite{Wampler2021,Ptaszynski2022} and multicyclic~\cite{Wampler2021} Markovian networks.
As broadly reviewed in~\cite{ICNF2023}, 
skewness of electrical current fluctuations was experimentally reported in tunnel junctions~\cite{Reulet2010,Fevrier2019}, avalanche diodes~ \cite{Gabelli2009}, quantum devices~\cite{Maisi2014} and metallic wire at cryogenic temperature~\cite{Pinsolle2018}.

In the present work, we derive a methodology to derive tight bounds on moments of entropy production. Besides recovering the generalized TUR~\cite{Falasco2020,Potts2019,hasegawa2019fluctuation} and Salazar's third-moment bound~\cite{Salazar2022}, it allows to find bounds on higher moments. As an illustration, we derive a novel, tight, bound between second and  third moment. 
We also write the tightest relations that hold close to equilibrium (in the limit of low entropy production). 

As the main result of this article, we find the relations that hold between the mean, variance and skewness of entropy production any white noise that is  thermodynamically consistent (in that it satisfies LDB). The bounds apply in particular to the flow (e.g., electric current) going through a purely dissipative device, in both equilibrium and far-from-equilibrium conditions.  
%
%
Our bound contains as particular cases several important special cases encountered in electronics, mechanics and chemistry: Johnson-Nyquist~\cite{Johnson1928,Nyquist1928} or Einstein diffusion process (Brownian motion~\cite{Einstein1905}); shot noise or any bidirectional Poisson process~\cite{Wyatt1999,Freitas2021stochastic,IFAC2023}.
Moreover, because any colored noise can be regarded as a white noise passing through a linear 
filter, we also introduce a corrected formula including the filtering effect of some system (e.g. experimental setup) through its impulse response.



\emph{Problem Statement.}---
For a real random observable $X$, let $m_k=\langle X^k \rangle$ be its $k$th moment, for 
$k=1,2,3,\ldots$
In this paper, $\langle \cdot \rangle$ always denotes the expectation operator, i.e. ensemble average.
The \emph{Hamburger moment problem}, in probability theory, consists in characterising all possible sequences $m_1,m_2,m_3,\ldots,m_k$ that indeed emerge as the $k$ first moments of some arbitrary random observable $X$. For instance for $k=2$, we find that real numbers  $m_1$ and $m_2$ are valid moments if and only if $m_2 \geq m_1^2$.  One may also state the problem for random observables of a certain type, for instance nonnegative random variables (Stieltjes problem). We find that real numbers $m_1$ and $m_2$ are moments of some nonnegative random observable if and only if $m_2 \geq m_1^2$ and $m_1\geq 0$. The extension for arbitrary $k$ is a standard topic of probability theory. 

In this article we ask
which real numbers can possibly arise as the moments (or, equivalently, cumulants) of entropy production of a (classical) system satisfying 
LDB, as defined below. 

We also analyse the near-equilibrium situation, and the case of white noise. We characterise all possible cumulants of order one (mean), two (variance) and three (skewness) for thermodynamically consistent white noise.




\emph{Consequences of the local detailed balance.}---Let $\Omega$ be a probability space with probability measure $p$ and an involution $\omega \mapsto \overline{\omega}$ (`involution' means that $\overline{\overline{\omega}}=\omega$). This defines $\overline{p}(A)=p(\overline{A})$ for any event $A\subseteq \Omega$.
Without loss of generality, and for the sake of simplicity of notations, we assume a discrete space, so that we can write, for an observable $f:\Omega \to \mathbb{R}$, the mean as $\langle f \rangle=\sum_{\omega \in \Omega} f(\omega) p(\omega)$ (even though in some cases the space $ \Omega$ is continuous, and this sum should be implicitly understood as an integral).

Let 
$\Dsigma \equiv\ln (p/\overline{p})$. 
We call this observable the `entropy production', in reference to the situations where this observable is indeed endowed with this physical meaning (up to Boltzmann's constant $k_B$). This includes the cases where $\Omega$ is the space of trajectories, over some time interval, of a Markov process modelling an overdamped physical process subject to a constant or time-symmetric protocol (the protocol referring to the transition probabilities characterizing the Markov process), and the involution is simply the time-reversal of the trajectory (i.e. the sequence of states traversed by the trajectory, read in reverse order), as a direct consequence of LDB~\cite{Falasco2020,Potts2019,hasegawa2019fluctuation,maes2021local,Peliti2021}. 
Notice however that our results hold mathematically for this observable whether or not it has the physical meaning of an entropy production. Indeed it may have very different meanings, such as the magnetization of a spin system at equilibrium if the involution is the spin reversal~\cite{Falasco2020,Guioth2016}.

Our question is to characterize all the possible values taken by the moments of the entropy production, $m_k=\langle \Dsigma^k \rangle$ for $k=1,2,3,\ldots$

First observe that all moments are nonnegative. If $k$ is odd,  we can write indeed
\begin{equation}
m_k=\sum_{\omega \in \Omega} p(\omega) \ln^k \frac{p(\omega)}{p(\overline{\omega})}= \sum_{\{\omega,\overline{\omega}\}}  (p(\omega)-p(\overline{\omega})) \ln^k \frac{p(\omega)}{p(\overline{\omega})} \geq 0,
\end{equation}
where the second sum runs over all unordered pairs $\{\omega,\overline{\omega}\}$ (counted once).
For even $k$ we have:
\begin{equation}
m_k= \sum_{\{\omega,\overline{\omega}\}}  (p(\omega)+p(\overline{\omega})) \ln^k \frac{p(\omega)}{p(\overline{\omega})} \geq 0.
\end{equation}

To go further, we first focus on the simplest case  $\Omega=\{\omega, \overline{\omega} \}$.  
Then $p(\omega)+p(\overline{\omega})=1$. We assume without loss of generality that $p(\omega)\geq 1/2$. We can write, for even $k$:
\begin{equation}
m_k=\ln^k \frac{p(\omega)}{p(\overline{\omega})}
\end{equation}
and for odd $k$ 
\begin{equation}
m_k= \ln^k \frac{p(\omega)}{p(\overline{\omega})} \frac{p(\omega)-p(\overline{\omega})}{p(\omega)+p(\overline{\omega})}
=\ln^k \frac{p(\omega)}{p(\overline{\omega})} \tanh \frac{1}{2}\ln \frac{p(\omega)}{p(\overline{\omega})}.
\end{equation}
Reparametrizing with $s=\ln \frac{p(\omega)}{p(\overline{\omega})}$ (for any $s\geq 0$), we see that the list of possible moments $(m_1,m_2,m_3,m_4,\ldots)$ in this simple case is $(s \tanh (s/2),s^2, s^3 \tanh (s/2), s^4, \ldots)$, for any $s\geq 0$.

Now consider a general $\Omega$, containing (possibly infinitely) many pairs $\{\omega, \overline{\omega} \}$. Then we can compute $m_k$ as a convex combination (weighted average) of the moments computed over each pair $\{\omega,\overline{\omega}\}$, weighted with probability $p(\{\omega,\overline{\omega}\})=p(\omega)+p(\overline{\omega})$:		
\begin{equation}
m_k=  \sum_{\{\omega,\overline{\omega}\}} p(\{\omega,\overline{\omega}\}) \frac{p(\omega)\pm p(\overline{\omega})}{p(\omega)+p(\overline{\omega})} \ln^k \frac{p(\omega)}{p(\overline{\omega})}
\end{equation}		 
Since the probability distribution over all pairs $\{\omega,\overline{\omega}\}$ can be arbitrary,  the set $S$ of  possible vector of moments $(m_1,m_2, m_3, m_4, m_5, \ldots)$ is the convex hull (i.e., set of convex combinations) of points of the curve $(s \tanh \frac{s}{2}, s^2,s^3 \tanh  \frac{s}{2}, s^4 ,s^5 \tanh  \frac{s}{2}, \ldots )$. This in itself is a complete, albeit implicit, characterization of the possible values taken by the moments (of all orders) of entropy production.

\begin{figure}[]
\psfragscanon
\psfrag{0}[cc][cc]{0}
\psfrag{m2}[cc][cc]{$m_2$}
\psfrag{m3}[cc][cc]{$m_3$}
\psfrag{Bound}[cl][cl]{\color{b}$\ds m_2^{3/2} \tanh \frac{ m_2^{1/2}}{2}$}
\includegraphics[scale=1]{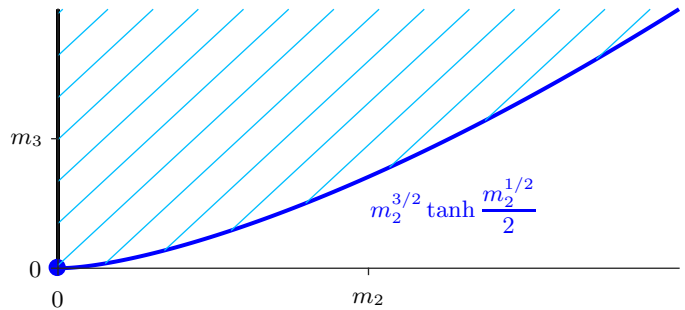}
\caption{
Illustration of the set of all possible pairs $(m_2, m_3)$ for entropy production.  The domain of validity (hatched in sky blue), defined by \eqref{eq:m2_m3}, is obtained as the convex hull of the blue boundary curve (which is thus included in the domain). The black boundary straight line $m_2 = 0$ (lower bound for $m_2$ in \eqref{eq:m2_m3}) is \emph{excluded}, save for the blue dot at the origin, which corresponds to the equilibrium (no entropy production).
}
\label{fig_PRL_2024_m2_m3}
\end{figure}

We are usually interested in moments of specific orders. As illustration, the set of all possible pairs $(m_2, m_3)$ for entropy production is the convex hull of the curve $(s^2, s^3 \tanh \frac{s}{2})$ 
(see \cref{fig_PRL_2024_m2_m3}) . It is thus characterized by 
\begin{equation}
\label{eq:m2_m3}
m_3 \geq m_2^{3/2} \tanh \frac{ m_2^{1/2}}{2}>0,
\end{equation}
(along with the trivial equilibrium point $m_2=0=m_3$). This means that every pair $(m_2,m_3)$ satisfying this relation is the second and third moments of the entropy production of some system, and conversely. The same methodology applied to $(m_1,m_2)$ recovers the generalized Thermodynamic Uncertainty Principle~\cite{Falasco2020,Potts2019,hasegawa2019fluctuation}
or to $(m_1,m_3)$ recovers~\cite{Salazar2022}.











\emph{Moments near equilibrium.}---We now consider the problem of characterizing moments near equilibrium, i.e., for small values of mean entropy production $m_1=\langle \Dsigma \rangle$. Remember that if $m_1=0$, then all the moments are zero (equilibrium, with $p\equiv \overline{p}$).

Consider once again the set $S$ of all possible moments $(m_1,m_2,m_3, \ldots)$. Recall that a cone is a subset of a vector space that is stable under positive linear combinations. Call $C$ the cone generated by the convex set $S$, which is the set of half-lines starting from the origin and meeting $S$. This cone is generated by positive linear combinations of points of the form  $(s \tanh (s/2),s^2,s^3 \tanh (s/2),s^4,\ldots)$. 

This cone is equivalently characterized by the convex set of all values taken by the `scaled coordinates' $(m_2/m_1,m_3/m_1, m_4/m_1, \ldots)$ (for all non-zero points in $S$) indicating the directions of the half-lines in $C$. In this representation, $C$ is  the convex hull of points $(m_2/m_1, m_3/m_1, m_4/m_1,\ldots)=(\frac{s}{\tanh (s/2)},s^2, \frac{s^3}{\tanh(s/2)}, \ldots)$. For the first three moments, we thus have
\begin{equation}
	\label{eq:main m2/m1}
	\ds
	2
	<
	\frac{m_2}{m_1}
	\leq
	\frac{\ds\sqrt{\frac{m_3}{m_1}}}{\ds\tanh\Big(\frac{1}{2}\sqrt{\frac{m_3}{m_1}}\Big)} \quad \text{and} \quad \frac{m_3}{m_1} > 0
\end{equation}
or
\begin{equation}
	\label{eq:main m2/m1 bis}
	\frac{m_2}{m_1}= 2  \quad \text{and} \quad m_3=0
\end{equation}
Notice that the r.h.s. of \eqref{eq:main m2/m1}  converges to $2$ for $m_3/m_1 \to 0$.
The (non-zero) moments $(m_1,m_2,m_3)$ of entropy production necessarily satisfy \eqref{eq:main m2/m1} or \eqref{eq:main m2/m1 bis}. This is a necessary condition, not sufficient in general, since the cone $C$ is larger than $S$. Nevertheless, intuitively, the cone $C$ is a good approximation of $S$ `near the origin'. We make this intuition precise.

Consider 
$C_\epsilon$ the (convex) set of points of the form $(m_2/\epsilon, m_3/\epsilon, \ldots)$, for all points  $(\epsilon, m_2,m_3,m_4 \ldots)$  in $S$. Thus $C_{\epsilon}$ encodes the `slice' of $S$ with coordinate $m_1=\epsilon$. By convexity of $S$ (which contains the origin),  the set $C_{\epsilon}$ is increasing with $\epsilon$: if $\epsilon < \epsilon'$ then $C_{\epsilon} \supset C_{\epsilon'}$. As $\epsilon \to 0$, then $C_{\epsilon}$ converges to $C$: every point in $C$ is eventually in $C_{\epsilon}$ for small enough \textcolor{h}{$\epsilon$}. Thus $C$, which is larger than $S$, can be seen as a good approximation of $S$ for those points with small enough coordinate $m_1 = \epsilon \approx 0$. 

Said otherwise, any numbers  $m_1,m_2,m_3$ satisfying   \eqref{eq:main m2/m1} or \eqref{eq:main m2/m1 bis} are possible moments $m_1,m_2,m_3$ of entropy production, provided that $m_1$ is small enough.

\emph{Moments and cumulants of white noise.}---The typical situation of interest where the bounds  \eqref{eq:main m2/m1} and \eqref{eq:main m2/m1 bis} apply is the entropy production of any overdamped stationary Markov process over any infinitesimal time interval, as $m_1=\mean{\Dsigma}$ vanishes with the time interval $\Delta t \to 0$. 

Let us examine a specific case of particular practical importance: the case of \emph{white} noise, which can be seen as a Markov process with a single state (no memory of the past).

A specific example is the electrical current flowing through an arbitrary dissipative electronic device,
like an homogeneous semiconductor or metallic bar
or a nonlinear diode or transistor,
subjected to a constant voltage 
$V$. 
The random current is adequately modelled as a white noise process~\cite{IFAC2023,ICNF2023}.
The total entropy production 
over a time interval $\Delta t$
reads
\begin{equation}
\label{eq:sigma}
\Dsigma = \frac{V}{\kBB T}\,\Dq
\end{equation}
where $\Dq$ is the random charge increment (integral of white noise current over $\Delta t$), $\kBB$ is Boltzmann's constant and $T$ is the constant temperature of the environment, 
assimilated to
a uniform thermal bath.
Thus, entropy production and charge increment over a same time interval are 
related by \eqref{eq:sigma}, proportional through a constant factor, the `thermodynamic force' $V/\kBB T$. Likewise, the electrical current $\dot{q}$ can be identified up to the same multiplicative constant to the entropy production rate $\dot{\sigma}$, both described by a white noise.

Similar situations occur in various context where a purely dissipative system subjected to a constant `thermodynamic force' (mechanical force, 
gradient of
chemical potential, of temperatures, of concentration, etc.) generates a random `flow' (speed, chemical flow, heat flow, matter flow, etc.).

Characterizing the moments of entropy production is thus characterizing the moments of electric current (speed, heat flow, etc.) in response to  a given external force, a problem of fundamental importance.

For any such white noise, the entropy production $\Dsigma$ over a time interval $\Delta t$ can be broken down as the sum of independent entropy productions over smaller intervals. Thus the cumulants $c_1, c_2, c_3, \ldots$ of entropy production over time $\Delta t$ are proportional to $\Delta t$, for any $\Delta t$ (small or large). Indeed the cumulants are additive for the sum of independent variables. Let us recall that the first cumulant $c_1$ is the mean (of entropy production over a time interval $\Delta t$), while $c_2$ is the variance and $c_3$ is the third central moment --- sometimes called skewness because it captures the non-symmetric nature of fluctuations on either side of the mean. The white noise is thus characterized by the cumulant rates $c_k/\Delta t$, which do not depend on $\Delta t$, and which we denote $\dot{c}_k$ (same unit as an inverse time). For instance $\dot{c}_1$ is the mean  entropy production rate.   

In the limit of short time intervals $\Delta t \to 0$, the cumulants $c_k$ and the moments $m_k$ coincide: for instance $c_2=m_2-m_1^2=m_2 + \mathcal{O}(\Delta t^2)$, while $c_2$ is proportional to $\Delta t$. Similarly, $c_3=m_3-3m_2m_1+ 2 m_1^3=m_3+ \mathcal{O}(\Delta t^2)$, etc. 

\begin{figure}[]
\psfragscanon
\psfrag{0}[cc][cc]{0}
\psfrag{2}[cc][cc]{2}
\psfrag{c3/c1}[cc][cc]{$c_3/c_1$}
\psfrag{c2/c1}[cc][cc]{$c_2/c_1$}
\psfrag{Bound}[tl][tl]{\color{b}\footnotesize$\ds \frac{\ds\sqrt{\frac{c_3}{c_1}}}{\ds\tanh\Big(\frac{1}{2}\sqrt{\frac{c_3}{c_1}}\Big)}$}
\includegraphics[scale=1]{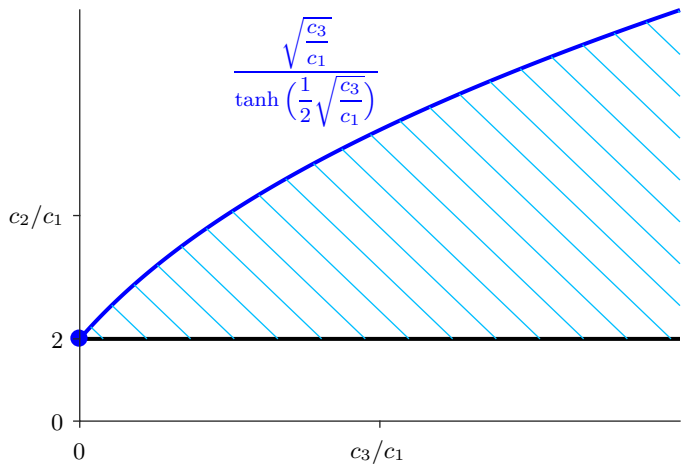}
\caption{
Possible values for the cumulants $c_1$ (mean), $c_2$ (variance), and $c_3$ (third central moment, indicating skewness) of thermodynamically consistent white noise.  The domain of validity (hatched in sky blue) defined by \cref{eq:main c2/c1,eq:main c2/c1 bis} is generated as the convex hull of the blue boundary curve (which is thus included in the domain). The black boundary straight line $c_2/c_1 = 2$ is \emph{excluded}. 
 Gaussian white noise is necessarily at the blue dot $(0,2)$, while the blue curve is otherwise populated by bidirectional Poisson processes. The same figure can be seen as representing the possible values of $m_3/m_1$ (horizontal axis) vs $m_2/m_1$ (vertical axis) for arbitrary moments $m_1,m_2,m_3$ of entropy production, as in \cref{eq:main m2/m1,eq:main m2/m1 bis}.
}
\label{fig_PRL_2024_bound_white_noise}
\end{figure}

Thus, the relations \eqref{eq:main m2/m1}\eqref{eq:main m2/m1 bis} hold for cumulants of white noise, which satisfy either
\begin{equation}
	\label{eq:main c2/c1}
	\ds
	2
	< 
	\frac{c_2}{c_1}
	\leq
	\frac{\ds\sqrt{\frac{c_3}{c_1}}}{\ds\tanh\Big(\frac{1}{2}\sqrt{\frac{c_3}{c_1}}\Big)} \quad \text{and} \quad \frac{c_3}{c_1} >0
\end{equation}
or
\begin{equation}
	\label{eq:main c2/c1 bis}
\frac{c_2}{c_1}= 2  \quad \text{and} \quad c_3=0
\end{equation}
In this equation, the ratios $c_2/c_1$ and $c_3/c_1$ do not depend on observation time $\Delta t$ and can be replaced (equivalently) with the rate ratios  $\dot{c}_2/\dot{c}_1$ and $\dot{c}_3/\dot{c}_1$.
See \cref{fig_PRL_2024_bound_white_noise} for illustration.

In fact, it characterizes the possible cumulants of thermodynamically consistent white noise, not only at  short times (where they coincide with moments), but also at arbitrarily long times (by time-proportionality of cumulants).

In particular Gaussian white noise, which has necessarily $c_3=0$ (no skewness, symmetric fluctuations around the mean), corresponds to the case \eqref{eq:main c2/c1 bis}, thus \emph{must} satisfy $c_2=2c_1$, which is a particular case of the \emph{fluctuation-dissipation theorem}~\cite{Kubo1966} expected in the near-equilibrium, linear response regime. This relation is also called Johnson-Nyquist's formula in the case of a linear electrical resistor or Einstein’s diffusion law~\cite{Einstein1905} in mechanics. Conversely, when the skewness is zero, one must obey the fluctuation-dissipation relation $c_2=2c_1$. In other words, fluctuations in a purely dissipative device that strictly exceed the fluctation-dissipation regime cannot be purely Gaussian and must 
exhibit some positive skewness.

Relation \eqref{eq:main c2/c1} holds with equality for a bidirectional Poisson random process, i.e., a white noise that is the sum of two Poisson processes with rates $\lambda_+$ and $\lambda_-$. Every arrival in the positive  (resp., negative) process generates an entropy of $\ln \frac{\lambda_+}{\lambda_-}$ (resp., the opposite). For instance in an electronic device, a charge carrier (electron, charge $q_e$) passing through the device subjected to a voltage $V$ generates an entropy  $\pm \ln \frac{\lambda_+}{\lambda_-}=\pm \frac{Vq_e}{kT}$. In this context, the bidirectional Poisson process is called \emph{shot noise}~\cite{IFAC2023,Wyatt1999}.

For such a bidirectional Poisson process, the first cumulants are the mean entropy production $c_1= (\lambda_+ - \lambda_-)  \ln \frac{\lambda_+}{\lambda_-} \Delta t$, the variance $c_2= (\lambda_+ + \lambda_-)  \ln^2  \frac{\lambda_+}{\lambda_-} \Delta t$, the skewness $c_3= (\lambda_+ - \lambda_-)  \ln^3 \frac{\lambda_+}{\lambda_-} \Delta t$). Direct check up shows that all these white noise processes satisfy the relation with equality, and populate the top curve of the domain in \cref{fig_PRL_2024_bound_white_noise}. The interior of the domain is obtained by positive linear combinations of different Gaussian or bidirectional Poisson processes. This is also in line with Lévy-Khintchine theorem~\cite{applebaum2009levy},
stating
that every (mathematical) white noise is decomposable into a (possibly continuous) sum of Poisson and Gaussian noises.

This confirms that the relation \eqref{eq:main c2/c1}\eqref{eq:main c2/c1 bis} is not only necessary but also sufficient to characterize the possible cumulants of thermodynamically consistent white noise. 

Finally, observe that the l.h.s inequality term of \eqref{eq:main c2/c1} is a particular case of the Thermodynamic Uncertainty Relation~\cite{Horowitz2020}. Thus the r.h.s inequality can be seen as a sort of converse of Thermodynamic Uncertainty Relation, as providing an \emph{upper} bound on the variance of the entropy production.

\emph{Application to Nonlinear Electronic Devices.}---Electrical current flowing through dissipative devices is an important instance of white noise process, as introduced above.
That the current fluctuations in nonlinear devices 
exhibit non-zero skewness has been previously highlighted experimentally and theoretically in~\cite{IFAC2023,ICNF2023}.

The cumulants of $\Dq$  can be related to those of $\Dsigma$ through the proportionality relationship \eqref{eq:sigma}, and thus satisfy, using
\eqref{eq:main c2/c1}\eqref{eq:main c2/c1 bis}, either
\begin{equation}
	\label{eq:main Dq}
	\ds
	2
	< \frac{V}{\kBB T} \frac{\mean{\Dq^2}}{\mean{\Dq}}
	\leq
	\frac{\ds\frac{V}{\kBB T} \sqrt{\frac{\mean{\Dq^3}}{\mean{\Dq}}}}{\ds\tanh\bigg(\frac{1}{2}\frac{V}{\kBB T}\sqrt{\frac{\mean{\Dq^3}}{\mean{\Dq}}}\bigg)} \quad \text{and} \quad \frac{\mean{\Dq^3}}{\mean{\Dq}}>0
\end{equation}
or
\begin{equation}
	\label{eq:main Dq bis}
\mean{\Dq^2} = 
	2 \kBB T \, \frac{\mean{\Dq}}{V}  \quad \text{and} \quad \mean{\Dq^3}=0 
\text{.}
\end{equation}
\Cref{eq:main Dq bis} is the Johnson-Nyquist formula~\cite{Johnson1928,Nyquist1928} (a particular case of the fluctuation-dissipation theorem), known to be valid at least for linear resistors.

Beyond this case, \eqref{eq:main Dq}
is valid for \emph{any} white noise physically generated within any nonlinear device.
From the knowledge of the mean and variance (varying with $V$), solving \eqref{eq:main Dq} for $\sqrt{\mean{\Dq^3}/\mean{\Dq}}$, provides a lower bound for the skewness of the current fluctuations, as a function of the applied voltage $V$ (possibly very large, allowing to investigate the far-from-equilibrium regime).


We aim at exploiting \eqref{eq:main Dq} to discuss the skewness of the current fluctuations in MOS (Metal-Oxide Semiconductor) (field-effect) transistors, key 
elements of modern integrated circuits in silicon.
A MOS transistor can be regarded as a structure with \emph{three} accesses: Gate, Source and Drain. 
In our analysis,  the gate-to-source voltage 
is assumed fixed, thus the transistor reduces to a nonlinear resistance between \emph{two} accesses, the source and the drain.
Within our notations, $\Dq/\Dt$ is the (drain-to-source) current, a white noise whose statistics depend on the constant source-to-drain voltage difference $V$.

The classical `long-channel' MOS transistor theory predicts, in strong inversion regime, the average current \cite{Tsividis2011}:
\begin{equation}
	\label{eq:MOS mean}
	\frac{\mean{\Dq}}{\Dt} = 
	\begin{cases}
		\ds \beta \, \Big(\Vsat V-\frac{V^2}{2}\Big) &\quad\text{if $V < \Vsat$} \\
		\ds \beta \, \frac{\Vsat^2}{2}
		&\quad\text{if $V \geq \Vsat$.}
	\end{cases}
\end{equation}
$\beta$ and saturation voltage $\Vsat$ are  constant which absorbs 
some transistor and physical parameters.
The variance of the white noise may be compactly written as \cite{Tsividis2011, Tedja1994}:
\begin{equation}
	\label{eq:MOS var}
\frac{\mean{\Dq^2}}{\Dt}
	= 2 \kBB T \, \beta \Vsat \, \frac{2}{3} \frac{1+\eta+\eta^2}{1+\eta}
\end{equation}
where 
\begin{equation}
	\label{eq:eta}
	\eta = 
	\begin{cases}
		\displaystyle 1 - \frac{V}{\Vsat} \quad \text{if } V \leq \Vsat \\
		\displaystyle 
		\mathrlap{0} \hphantom{1 - \frac{V}{\Vsat}} \quad \text{if } V \geq \Vsat
		\text{.}
	\end{cases}
\end{equation}
As already reported in~\cite{IFAC2023}, following these equations it is straighforward to check that  
$\mean{\Dq^2}/\mean{\Dq}$
ranges from 2 (thus reaching Johnson-Nyquist's lower bound  of \eqref{eq:main Dq bis}) for $V \to 0$ to $8/3$ for $V \geq \Vsat$.
This result is consistent with the fact that, for $V \to 0$, the two-access transistor behaves like a linear resistor.

\begin{figure}
\centering
\psfragscanon
\psfrag{V/Vsat}[cc][cc]{$V/\Vsat$}
\psfrag{sqrt(<q3>/<q>)/qe}[cc][cc]{$\sqrt{\mean{\Dq^3}/\mean{\Dq}} \ [\qe]$}
\psfrag{0}[cc][cc]{$0$}
\psfrag{p/V}[cr][tr]{$\ds\frac{\kB T}{\qe\Vsat}$}
\psfrag{1}[cc][cc]{$1$}
\psfrag{Shot noise}[cl][cl]{\color{r}Shot noise (e.g. junction)}
\psfrag{MOS}[cl][cl]{\color{b}MOS transistor}
\psfrag{JN}[cl][cl]{\color{k}Gaussian (linear resistor)}
\includegraphics[scale=1]{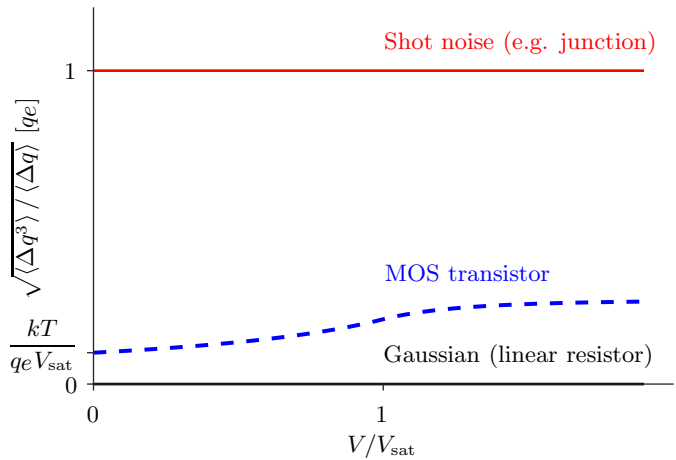}
\caption{
Minimum skewness (third central moment) of current through a MOS transistor, predicted from \eqref{eq:main Dq} (dashed curve). 
Gaussian and shot noise skewness are also depicted for comparison. 
}
\label{fig_PRL_2024_MOS}
\end{figure}

Thus for $V>0$, Johnson-Nyquist's prediction \eqref{eq:main Dq} is exceeded and 
\eqref{eq:main Dq bis} must instead apply. This allows to extract (numerically) a lower bound for the skewness as a function of $V$, see dashed line in \Cref{fig_PRL_2024_MOS}. The initial value can be computed analytically by a Taylor approximation of the tanh in \eqref{eq:main Dq bis}:
\begin{equation}
\sqrt{\mean{\Dq^3}/\mean{\Dq}}
\geq
\frac{\kB T/\qe}{\Vsat} \qe
\text{.}
\end{equation}

This value must be compared to the zero skewness of a pure Gaussian
noise, and to $\qe$ analytically computed for a shot noise (a bidirectional Poisson noise of intensity $q_e$, modelling the transfer of one electron at a time through the resistance), also measured experimentally in a tunnel junction~\cite{Reulet2010}.

This theoretical result proves that the white noise in a MOS transistor has a positive skewness and hence is not rigorously Gaussian, although this convenient assumption is widely used for noise modelling and circuit simulations~\cite{SSE2023}.

We find instructive to consider realistic numerical values for the parameters involved.
The order of magnitude of $\sqrt{\mean{\Dq^3}/\mean{\Dq}}$, in number of $\qe$, is determined by the $\kBB T/(\qe\Vsat)$.
At room temperature $\kBB T/\qe \approx \SI{25}{\milli\volt}$.
Regarding $\Vsat$, it typically ranges from a few $\si{\volt}$ in old $\si{\micro\meter}$ CMOS technologies and down to several hundreds of $\si{\milli\volt}$ in the most advanced decananometer technologies.
$\Vsat = \SI{250}{\milli\volt}$, as used in \cref{fig_PRL_2024_MOS}, results in $\sim \qe/10$.

\emph{Discussion.}---A natural continuation for those results would be the characterization of higher-order temporal correlations of non-white noise as produced by general Markov chains, in view to complement for example the recent results of autocorrelations and power spectral correlations\cite{dechant2023thermodynamic,dechant2023thermodynamic2,ohga2023thermodynamic,vanvu2023dissipation}.

\emph{Acknowledgement.}---This research was funded by Research Project ‘‘Thermodynamics of Circuits for Computation’’ of the National Fund for Scientific Research
(F.R.S.-FNRS) of Belgium.

\bibliographystyle{apsrev4-1} 
\bibliography{bib}

\end{document}